\documentclass[aps,pre, onecolumn,showpacs,12pt,preprintnumbers,showkeys]{revtex4}
\usepackage{amsfonts}

\usepackage{graphicx}
\usepackage{dcolumn}
\usepackage{bm}
\include{biblio}
\begin{document}

\title{Influence of asymmetric depletion of solvents on the electric double layer of  charged objects in binary polar solvent mixture}

\author{Jun-Sik Sin}
\email{js.sin@ryongnamsan.edu.kp}
\affiliation{Department of Physics, \textbf{Kim Il Sung} University, Taesong District, Pyongyang, DPR Korea}
\author{Chung-Sik Sin}
\affiliation{Department of Physics, \textbf{Kim Il Sung} University, Taesong District, Pyongyang, DPR Korea}

\begin{abstract}

 For binary solvent mixtures composed of ions and two kinds of polar solvents, electric double layer near a charged object is strongly affected by not only the binary solvent composition but also nature of solvents such as volume and dipole moment of a solvent molecule. Accounting for difference in sizes of solvents and orientational ordering of solvent dipoles, we theoretically obtain general expressions for the spatial distribution functions of solvents and ions, in planar geometry and within the mean-field approach. Although focusing on long-range electrostatic interaction, we neglect short-range interactions such as preferential solvation, our approach predicts an asymmetric depletion of the two solvents from the charged surface and a behavior of decreased permittivity of the binary solvent mixture. Furthermore, we suggest that the key factor for the depletion is the ratio of the solvent dipole moment to the solvent volume. The influence of binary solvent composition, volume of solvent and dipole moment of solvent on the number density of solvents, permittivity and differential capacitance are presented and discussed, respectively. We conclude that accounting for difference in the volume and dipole moment between polar solvents  is necessary for new approach to represent  more realistic situations such as preferential solvation.
\end{abstract}

\maketitle

\section{Introduction}

The behavior of binary solvent mixtures in the presence of dissolved ions and external electric fields is relevant to material scientific and biological applications such as manipulation of microfluids, protein stability and conformational changes \cite%
{Squires_rmp_2005, Whiteside_nature_2006, Psaltis_nature_2006, Timasheff_arbbs_1993, Davis_arbbs_2001, Record_APC_1998, Harries_JMCB_2008}. 
Recent experiments demonstrated that  the addition of another polar solvent to the aqueous solution can allow a DNA molecule to transit from elongated coil to compact globule, suggesting that the interaction between the DNA segments is strongly affected by the presence of this added solvent \cite%
{Baigl_biophys_2005}. It is well known that adding an excess amount of ethylalcohol to the aqueous solution counteracts the repulsion between charged DNA strands and precipitates DNA \cite%
{Parsegian_pnas_2000, Hultgren_biochem_2004, Stanely_biophys_2006}.

	Although  the properties of binary solvent mixtures such as surface tension and osmotic pressure have been theoretically investigated \cite%
{Tsori_pnas_2007, Onuki_jchemphys_2004, Onuki_jchemphys_2008, Andelman_jphyschemb_2009, Budkov_jchemphys_2016}, most of the theoretical models did not account for orientational ordering of solvent dipoles and non-uniform size effects of solvents.
     For aqueous electrolyte solution, authors of \cite%
{Iglic_bioelechem_2010, Gongadze_bioelechem_2012, Gongadze_electrochimica_2013}
presented a formula for the spatial dependence of the relative permittivity of an electrolyte near a highly charged surface with a modified Poisson-Boltzmann approach. Taking into account the mutual influence of the water molecules by means of the cavity field, the formula represented orientational ordering of water dipoles without non-uniform size effect of ions and solvents.
	The model of \cite%
{Gongadze_bioelechem_2012}
 was generalized by \cite%
{Gongadze_electrochimica_2015}, which considered the different size of positive and negative ions and water molecules in electric double layer, and obtained analytical expressions for ion spatial distribution functions. 

Recently, we modified  the  mean-field approach  including the different size effect of ions and water molecules and the orientational ordering of water dipoles presented in \cite%
{Sin_electrochimica_2015} in  order to better  describe counterion stratification \cite%
{Sin_electrochimica_2016}. Unlike previous Poisson-Boltzmann approaches \cite%
{Gouy_physfrance_1910,  Chapman_philos_1913, Bikerman_philmag_1942, Wicke_zelecchem_1952, Iglic_physfran_1996, Borukhov_prl_1997, Borukhov_electrochimica_2000, Bohnic_electrochimica_2001, Bohnic_bioelechem_2005, Chu_biophys_2007, Tresset_pre_2008, Kornyshev_physchem_2007, Biesheuvel_jcolloid_2007, Li_pre_2011, Li_pre_2012, Boschitsch_jcomchem_2012,  Popovic_pre_2013} , the approach of \cite%
{Sin_electrochimica_2016} has the ability to predict both counterion stratification \cite%
{Shapovalov_jphyschem_2006, Shapovalov_jphyschem_2007}  for mixtures of counterions of different species and decreased permittivity \cite%
{Onsager_amchem_1936, Kirkwood_chemphys_1939,  Booth_chemphys_1951, Booth_chemphys_1955, Butt_2003} of electrolyte solution near a charged surface. However, the approach can not be applied to cases of binary solvent mixtures. 

	In this paper, we will extend the approach of \cite%
{Sin_electrochimica_2016} to also treat binary solvent mixtures.  We will introduce a lattice statistics \cite%
{Bohnic_bioelechem_2005} where more than one cell can be occupied by each ion as in \cite%
{Popovic_pre_2013, Gongadze_electrochimica_2015} and also by each solvent molecule for considering effects of different sizes of ions and solvent molecules \cite%
{Gongadze_electrochimica_2015, Sin_electrochimica_2015, Sin_pccp_2016, Sin_electrochimica_2016}. Like our previous approach \cite%
{Sin_electrochimica_2016}, the present approach yields the corresponding expressions for ion and solvent spatial distribution functions.  In the section of results and discussion, the solvent distribution functions, the portion of water molecules to solvent molecules, permittivity of solvent mixtures and differential capacitance are estimated as functions of the binary solvent composition, volume and dipole moment of added solvent, respectively. Finally, asymmetric depletion of the two solvents from the charged surface is analyzed and the key factor for the depletion is suggested.

\section{Theory}

We consider a binary solvent mixture composed of ions of different species, water molecules and added polar solvent molecules near a charged planar surface, where an ion has charge $z_{i}e_{0}$ .
For the sake of simplicity, we have used a planar geometry for the charged objects, but the extension to cylindrical or spherical geometry is straightforward. 
Although preferential solvation plays important role for determining some properties of binary solvent mixtures, we assume that short-range interactions including preferential solvation are negligible. It is for this reason that a deep understanding of long-range interaction in binary solvent mixtures allows one to know not only the role of the interaction but also the contribution from short-range interactions relative to the long-range interaction.
The total free energy $F$ can be written in terms of the local electrostatic potential  $\psi\left(r\right)$ and the number densities of different ionic species $n_{i}\left(r\right)$  $\left(i=1\ldots m\right) $, the number density of water molecules
$n_{w}\left(r\right)=\left<\rho_{w}\left(\omega, r \right)\right>_{\omega}$  and the number density of added solvent molecules  $n_{a}\left(r\right)=\left<\rho_{a}\left(\omega, r \right)\right>_{\omega}$
 \begin{eqnarray}
	F=\int{d{\bf r}}\left(-\frac{\varepsilon_{0}\varepsilon E^2}{2}+e_{0}\psi\sum^{m}_{i=1}z_{i}n_{i}+ \left<\rho_{w}\left(\omega\right)\gamma{p_{w}}E\cos\left(\omega\right)\right>_{\omega}+\left<\rho_{a}\left(\omega\right)\gamma{p_{a}}E\cos\left(\omega\right)\right>_{\omega}\right)-\nonumber\\ \int{d{\bf r}}\left(\sum^{m}_{i=1}\mu_{i}n_{i}+\left<\mu_{w}\left(\omega\right)\rho_{w}\left(\omega\right)\right>_{\omega}+\left<\mu_{a}\left(\omega\right)\rho_a\left(\omega\right)\right>_{\omega}+Ts\right),
\label{eq:1}
\end{eqnarray}

where $\left<f\left(\omega\right)\right>_{\omega}={\int f\left(\omega\right)2\pi\sin\left(\omega\right)  d\omega}/\left(4 \pi\right)$ in which $\omega$ is the angle between the vector ${\bf p}$ and the normal to the charged surface.  Here ${\bf p}_{w}$ and ${\bf p}_{a}$ are the dipole moments of water molecules and added solvent molecules, respectively. {\bf E} stands for the electric field strength, while $z_{i}\left(i=1\ldots m\right)$ is the ionic valence of ions. The first term represents the self energy of the electrostatic field, where $\varepsilon$ is equal to $n^2$ , where  $n$ is the effective refractive index of binary solvent mixture. We assume that because the refractive indices of two solvents are slightly different from each other, the effective index of binary solvent mixture is taken as the value of water. The second one represents the electrostatic energy of the ions in the binary solvent mixture, where $e_{0}$ is the elementary charge. The third and fourth terms correspond to the electrostatic energy of water dipoles and added solvent molecules \cite% 
{Gongadze_bioelechem_2012}, where  $\gamma=\left(2+n^2\right)/2$,  $p_{w}=\left|{\bf p}_{w}\right|$,  $p_{a}=\left|{\bf p}_{a}\right|$ and $E=\left|{\bf E}\right|$. The next three terms stand for coupling of the system to a bulk reservoir, where $\mu_{i}$$\left(i=1\ldots m\right) $ are the chemical potentials of ions and $\mu_{w}\left(\omega\right)$ and  $\mu_{a}\left(\omega\right)$ are the chemical potentials of water dipoles and added solvent molecules with orientational angle $\omega$. $T$ and $s$ are the temperature and the entropy density, respectively.

Let's consider a unit volume of the binary solvent mixture near a charged plane. The entropy density is taken as the logarithm of the number of translational and orientational arrangements of non-interacting   $n_{i}\left(i=1\cdots m\right)$ ions,  $\rho_{w}\left(\omega_{i}\right)\Delta\Omega_{i}\left(i=1 \cdots k\right)$ water molecules and $\rho_{a}\left(\omega_{i}\right)\Delta\Omega_{i}\left(i=1 \cdots k\right)$ added solvent molecules where $\Delta\Omega_{i}=2\pi  \sin\left(\omega_{i}\right) \Delta\omega/\left(4\pi\right)$  is an element of a solid angle and $\Delta\omega=\pi/ k$. An ion of {\it i}th species, a water molecule and an added solvent molecule occupy volumes of $V_{i}\left(i=1\cdots m\right)$, $V_{w}$ and $V_{a}$, respectively.

Within the lattice statistics approach each particle in the binary solvent mixture occupies more than one cell of a lattice as in \cite%
 {Boschitsch_jcomchem_2012}. As in \cite%
 {Gongadze_electrochimica_2015}, orientational ordering of solvent dipoles as well as translational arrangements of ions is explicitly considered. 
In the same way as in \cite%
{Boschitsch_jcomchem_2012}, we first place $n_{i}$ ions of the volume $V_{i}$  in the lattice. Accounting for the orientational ordering of solvent dipoles, we first put in  $\rho_{w}\left(\omega_{i}\right) \left(i=0,1,...\right)$ water molecules of the volume $V_{w}$  and then $\rho_{a}\left(\omega_{i}\right) \left(i=0,1,...\right)$ added solvent molecules of the volume $V_{a}$ in the lattice. The number of arrangements $W$ is written as
 \begin{equation}
	W=\frac{N!}{\prod^{m}_{i=1}n_{i}!\cdot\lim_{k\rightarrow\infty}\prod^{k}_{i=1}\rho_{w}\left(\omega_{i}\right)\Delta\Omega_{i}!\cdot\lim_{k\rightarrow\infty}\prod^{k}_{i=1}\rho_{a}\left(\omega_{i}\right)\Delta\Omega_{i}!},
\label{eq:2}
\end{equation}
where
 \begin{equation}
  N=\sum^{m}_{i=1}n_{i}+\lim_{k\rightarrow\infty}\sum^{k}_{i=1}\rho_{w}\left(\omega_{i}\right)\Delta\Omega_{i}+\lim_{k\rightarrow\infty}\sum^{k}_{i=1}\rho_{a}\left(\omega_{i}\right)\Delta\Omega_{i}.
\label{eq:3}
\end{equation}

Rearranging the logarithms of factorials according to the Stirling formula, the expression for the entropy density, $s=k_{B}\ln W$, is obtained as follows 
 \begin{eqnarray}
	\frac{s}{k_{B}}=\ln W=N\ln N-\sum^{m}_{i=1}n_{i}\ln n_{i}-\lim_{k\rightarrow\infty}\sum^{k}_{i=1}\rho_{w}\left(\omega_{i}\right)\Delta\Omega_{i} \ln\left[\rho_{w}\left(\omega_{i}\right)\Delta\Omega_{i}\right]-\nonumber\\
\lim_{k\rightarrow\infty}\sum^{k}_{i=1}\rho_{a}\left(\omega_{i}\right)\Delta\Omega_{i} \ln\left[\rho_{a}\left(\omega_{i}\right)\Delta\Omega_{i}\right],
\label{eq:4}
\end{eqnarray}
where $k_{B}$  is the Boltzmann constant.

This expression is consistent with the fact that in the physical point of view, the entropy density should be  symmetric in different species of ions.

Either ions or solvent molecules occupy all lattice cells\cite%
{Iglic_physfran_1996, Li_pre_2012, Li_pre_2011, Boschitsch_jcomchem_2012, Gongadze_bioelechem_2012}, therefore
 \begin{eqnarray}
	1=\sum^{m}_{i=1}n_{i}V_{i}+n_{w}V_{w}+n_{a}V_{a}.
\label{eq:5}
\end{eqnarray}

Differences in size of ions and solvent molecules are taken into account by  Eq.(\ref{eq:5}), which means  that single ions and single solvent  molecules  occupy different number of  lattice sites.
In fact, in order to account for different size of particles exactly, Eq.(\ref{eq:2}) should contain the difference in size of particles as shown in \cite%
{Popovic_pre_2013}. Although Eq.(\ref{eq:2}) is a good approximation for dilute solutions everywhere in the system, but not in the close vicinity of the charged surface. We emphasize that for the analytical tractability of resultant expressions we use the entropy  formula, Eq.(\ref{eq:2}).

Applying the method of undetermined multipliers, the Lagrangian of the binary solvent mixture is 
 \begin{eqnarray}
	L=F-\int\lambda\left({\bf r}\right)\left(1-\sum^{m}_{i=1}n_{i}V_{i}-n_{w}V_{w}-n_{a}V_{a}\right)d{\bf r},
\label{eq:6}
\end{eqnarray}

where $\lambda$ means a local Lagrange parameter.
The Euler$-$Lagrange equations should be obtained and solved with respect to the functions $n_{i}$, $\rho_{w}\left(\omega\right)$ and $\rho_{a}\left(\omega\right)$.
The variations of the Lagrangian with respect to $n_{i}$, $\rho_{w}\left(\omega\right)$ and  $\rho_{a}\left(\omega\right)$  yield equations related to $n_{i}$ , $\rho_{w}\left(\omega\right)$ and $\rho_{a}\left(\omega\right)$  in the binary solvent mixture:
 \begin{eqnarray}
	\frac{\delta L}{\delta n_{i}}=e_{0}z_{i}\psi-\mu_{i}+k_{B}T\ln\left(n_{i}/N\right)+V_{i}\lambda =0,   \left(i=1\cdots m\right),
\label{eq:7}
\end{eqnarray}
 \begin{eqnarray}
	\frac{\delta L}{\delta \rho_{w}\left(\omega\right)}=\gamma p_{w}\beta E\cos\left(\omega\right)-\mu_{w}\left(\omega\right)+k_{B}T\ln\left(\rho_{w}\left(\omega\right)\Delta\Omega/N\right)+V_{w}\lambda =0.
\label{eq:8}
\end{eqnarray}
\begin{eqnarray}
	\frac{\delta L}{\delta \rho_{a}\left(\omega\right)}=\gamma p_{a}\beta E\cos\left(\omega\right)-\mu_{a}\left(\omega\right)+k_{B}T\ln\left(\rho_{a}\left(\omega\right)\Delta\Omega/N\right)+V_{a}\lambda =0.
\label{eq:81}
\end{eqnarray}
The first boundary condition is $\psi\left(r\rightarrow\infty\right)=0$ which represents the fact that the origin of the electric potential is placed at $r\rightarrow\infty$. The boundary conditions for other physical quantities are  $n_{i}\left(r\rightarrow\infty\right)=n_{ib}$,  $n_{w}\left(r\rightarrow\infty\right)=n_{wb}$,  $n_{a}\left(r\rightarrow\infty\right)=n_{ab}$ and   $\lambda\left(r\rightarrow\infty\right)=\lambda_{b}$, where  $n_{ib}$, $n_{wb}$, $n_{ab}$   and $\lambda_{b}$  represent the ionic concentration,  the number density of water molecules, the number density of added solvent molecules and the Lagrange parameter at  $r\rightarrow\infty$, respectively. 

With the help of the boundary conditions,  we get the following equations from Eq.(\ref{eq:7}), (\ref{eq:8})
 \begin{eqnarray}
	\frac{n_{i}}{N}=\frac{n_{ib}}{N_{b}}\exp\left(-hV_{i}-e_{0}z_{i}\psi\right),    \left(i=1\cdots m\right).
\label{eq:9}
\end{eqnarray}

\begin{eqnarray}
	\frac{n_{w}}{N}=\frac{n_{wb}}{N_{b}}\exp\left(-h V_{w}\right)\langle\exp\left(-\gamma p_{w}\beta E\cos\left(\omega\right)\right)\rangle_{\omega}.
\label{eq:10}
\end{eqnarray}

\begin{eqnarray}
	\frac{n_{a}}{N}=\frac{n_{ab}}{N_{b}}\exp\left(-h V_{a}\right)\langle\exp\left(-\gamma p_{a}\beta E\cos\left(\omega\right)\right)\rangle_{\omega}.
\label{eq:101}
\end{eqnarray}
where $h\equiv\lambda-\lambda_{b}$ and \cite%
{Iglic_bioelechem_2010, Gongadze_bioelechem_2012}
\begin{eqnarray}
	\left\langle\exp\left(-\gamma p_{w}\beta E\cos\left(\omega\right)\right)\right\rangle_{\omega}=\frac{2 \pi\int^{0}_{\pi}d\left(cos\left(\omega\right)\right)\exp\left(-\gamma p_{w}\beta E\cos\left(\omega\right)\right)}{4\pi}=\frac{\sinh\left(\gamma p_{w} E \beta\right)}{\gamma p_{w} E \beta}.
\label{eq:11}
\end{eqnarray}
With the help of Eqs. (\ref{eq:5}),(\ref{eq:3}), summing  Eqs. (\ref{eq:9}),(\ref{eq:10},(\ref{eq:101}) over all species of particles results in the following equations for the number densities of ions, water molecules and added solvent molecules.
\begin{equation}
	n_{i}=\frac{n_{ib}\exp\left(-V_{i}h-e_{0}z_{i}\beta\psi\right)}{D},    \left(i=1\cdots m\right),
\label{eq:14}
\end{equation}

\begin{equation}
	n_{w}=\frac{n_{wb}\exp\left(-V_{w}h\right)\langle\exp\left(-\gamma p_{w}\beta E\cos\left(\omega\right)\right)\rangle_{\omega}}{D},
\label{eq:15}
\end{equation}

\begin{equation}
	n_{a}=\frac{n_{ab}\exp\left(-V_{a}h\right)\langle\exp\left(-\gamma p_{a}\beta E\cos\left(\omega\right)\right)\rangle_{\omega}}{D},
\label{eq:151}
\end{equation}

\begin{eqnarray}
	\sum^{m}_{i=1}n_{ib}\left(\exp\left(-V_{i}h-e_{0}z_{i}\beta\psi\right)-1\right)+n_{wb}\left(\exp\left(-V_{w}h\right)\frac{\sinh\left(\gamma p_{w}\beta E\right)}{\gamma p_{w}\beta E}-1\right)+\nonumber\\
n_{ab}\left(\exp\left(-V_{a}h\right)\frac{\sinh\left(\gamma p_{a}\beta E\right)}{\gamma p_{a}\beta E}-1\right)=0.
\label{eq:16}
\end{eqnarray}
where $D=\sum^{m}_{k=1} V_{k}n_{kb}\exp\left(-V_{k}h-e_{0}z_{k}\beta\psi\right)+V_{w}n_{wb}\exp\left(-V_{w}h\right)\left\langle\exp\left(-\gamma p_{w}\beta E\cos\left(\omega\right)\right)\right\rangle_{\omega}+V_{a}n_{ab}\exp\left(-V_{a}h\right)\left\langle\exp\left(-\gamma p_{a}\beta E\cos\left(\omega\right)\right)\right\rangle_{\omega}$.

It can be certified that the result found here includes all the previous ones \cite%
{Gongadze_bioelechem_2012, Boschitsch_jcomchem_2012, Gongadze_electrochimica_2015, Sin_electrochimica_2016} as special cases. 

The Euler$-$Lagrange equation for  $\psi\left(r\right)$  yields the Poisson equation
\begin{eqnarray}
	\nabla\left(\varepsilon_{0}\varepsilon_{r}\nabla\psi\right)=-e_{0}\sum^{m}_{i=1}z_{i}n_{i},	
\label{eq:18}
\end{eqnarray}
where
 \begin{eqnarray}
	\varepsilon_{r} \equiv n^2+\frac{|{\bf P}|}{\varepsilon_{0}{E}}.
\label{eq:19}
\end{eqnarray}
${\bf P}$ is the total polarization vector due to orientational ordering of solvent dipoles \cite%
{Gongadze_bioelechem_2012}. 
The planar symmetry of this problem results in the fact that the electric field strength is perpendicular to the charged surface and have an equal magnitude at all points equidistant from the charged surface. The $x$ axis is perpendicular to the charged surface and points in the direction of the bulk solution. As a consequence, {\bf E} and {\bf P}  have only an $x$ component and {\bf P}  is taken as \cite%
{Gongadze_bioelechem_2012}
\begin{eqnarray}
	{\bf P}\left(x\right)=\left(n_{w}\left(x\right)\left(\frac{2+n^2}{3}\right)p_{w}\mathcal{L}\left(\gamma{p_{w}}E\beta \right)+n_{a}\left(x\right)\left(\frac{2+n^2}{3}\right)p_{a}\mathcal{L}\left(\gamma{p_{a}}E\beta \right)\right)\hat{{\bf e}}, 
\label{eq:20}
\end{eqnarray}
where a function $\mathcal{L}\left(u\right)=\coth\left(u\right)-1/u$ is the Langevin function, $\hat{{\bf e}}={{\bf E}/E}$ and $\beta=1/\left(k_{B}T\right)$ .

The electrostatic potential and number densities of ions, water molecules and added solvent molecules are determined by solving Eqs. (\ref{eq:14}),(\ref{eq:15}),(\ref{eq:151}),(\ref{eq:16}),(\ref{eq:18}).

\section{Results and Discussion}
With the help of the boundary conditions $\psi\left(x\rightarrow\infty\right)=0$  and $E\left(x=0\right)=\sigma/\left(\varepsilon_{0}\varepsilon_{r}\left(x=0\right)\right)$, we solve coupled - differential equations Eqs. (\ref{eq:14}),(\ref{eq:15}),(\ref{eq:151}),(\ref{eq:16}),(\ref{eq:18})  for $n_{i}\left(i=1\ldots m\right), n_w, n_a, \psi$ by using the fourth order Runge-Kutta method. For all the calculations, the temperature $T$ and the bulk ionic concentration of binary solvent mixture have been taken equal to $298.15K$ and  $0.01mol/l$, respectively. In calculations, the ionic valence of all ions is $Z_1=-Z_2=+1$.  As in \cite%
{Gongadze_bioelechem_2012, Gongadze_electrochimica_2015}, the water dipole moment should be $3.1D$ so that far away from the charged surface, $x=\infty$, the relative permittivity of the electrolyte reaches the value of pure water.  In the same way, the dipole moment of an ethylalcohol molecule is chosen as 2.77D using the permittivity of ethyalcohol, 20 \cite%
{Andelman_jphyschemb_2009, Butt_2003}.

\begin{figure}
\begin{center}
\includegraphics[width=1\textwidth]{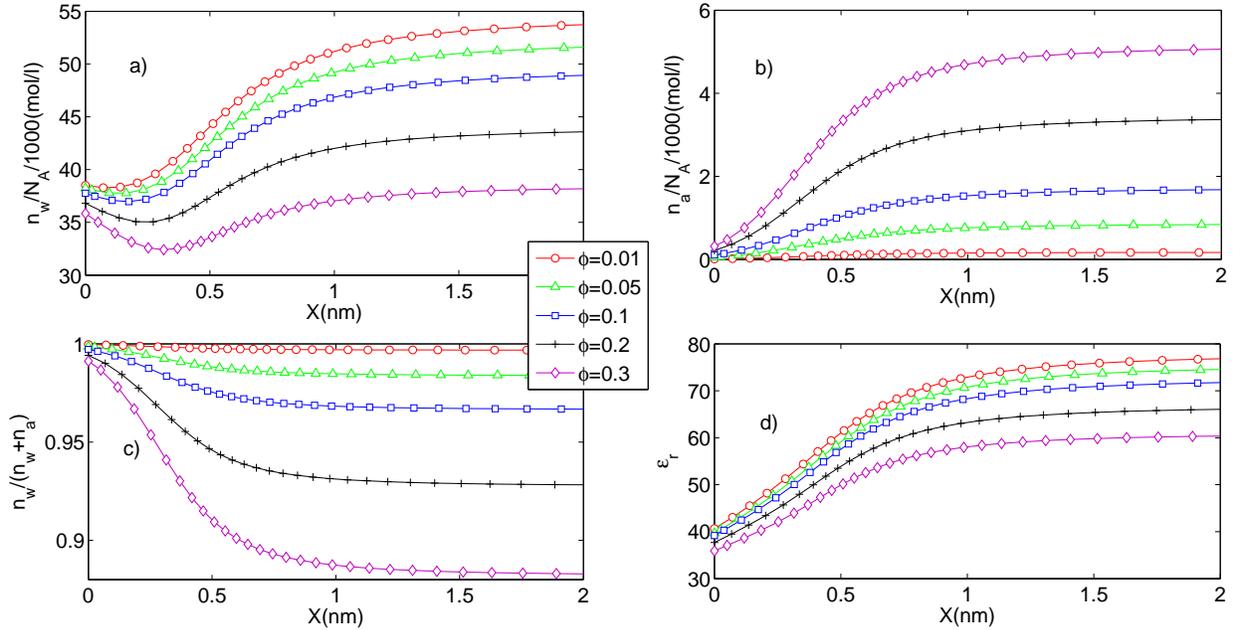}
\caption{(Color online) The number density of water molecules (a), the number density of ethylalcohol molecules (b), the portion of water molecules in solvent mixture (c) and the permittivity (d) as a function of the distance from the charged surface in binary solvent mixtures. The volumes of ions are $V_{+}=0.1nm^3, V_{-}=0.2nm^3$, respectively. The surface charge density of the charged surface are $\sigma= 0.2C/m^2$}
\label{fig:1}
\end{center}
\end{figure}

Fig. \ref{fig:1}(a) shows the spatial dependence of the number densities of water molecules near a charged surface with surface charge density  $\sigma=0.2C/m^2$. Circles, Triangles, Squares, Plus Signs and Diamonds stand for the number densities of water molecules for the cases having the binary solvent composition of $ \phi=0.01, 0.05, 0.1, 0.2, 0.3$  within our approach, respectively.  The binary solvent composition is defined as the portion of volume occupying added solvent in binary solvent mixture. Fig. \ref{fig:1}(a) represents that due to the accumulation of counterions, the water molecules are partially depleted from the region near the charged surface and a valley of the number density of water molecules is formed. It is shown that the larger the binary solvent composition, the lower the number density of water molecules everywhere near the charged surface.

Fig. \ref{fig:1}(b) displays the number densities of ethylalcohol molecules near a charged surface with the surface charge density $\sigma=0.2C/m^2$.  Fig. \ref{fig:1}(b) indicates that contrary to the number density of water molecules, the number densities of ethylalcohol molecules monotonously decrease with decreasing the distance from the charged surface for all cases. This means that if one of two solvents in binary solvent mixture has large volume and small dipole moment compared to other solvent, the first solvent is expelled more strongly near the charged surface than for the second solvent. This fact is clearly demonstrated from Eqs.(\ref{eq:10}, \ref{eq:11}).

Fig. \ref{fig:1}(c) shows the portion of water molecules to both solvent molecules according to the distance from the charged surface.  Obviously, the portion of water molecules increases with decreasing the distance from the charged surface. In other words, one might expect an increase in exclusion of the solvent having the lower permeability and larger volume near the wall.  This fact is due to the same reason as in the Fig. \ref{fig:1}(b).

Fig. \ref{fig:1}(d) displays the spatial dependence of relative permittivity according to the distance from the charged surface. For the case of every binary solvent composition, the permittivity profile monotonously decreases with decreasing the distance from the charged surface. It is also shown that the lower the binary solvent composition, the higher the permittivity of binary solvent mixtures.

\begin{figure}
\includegraphics[width=1\textwidth]{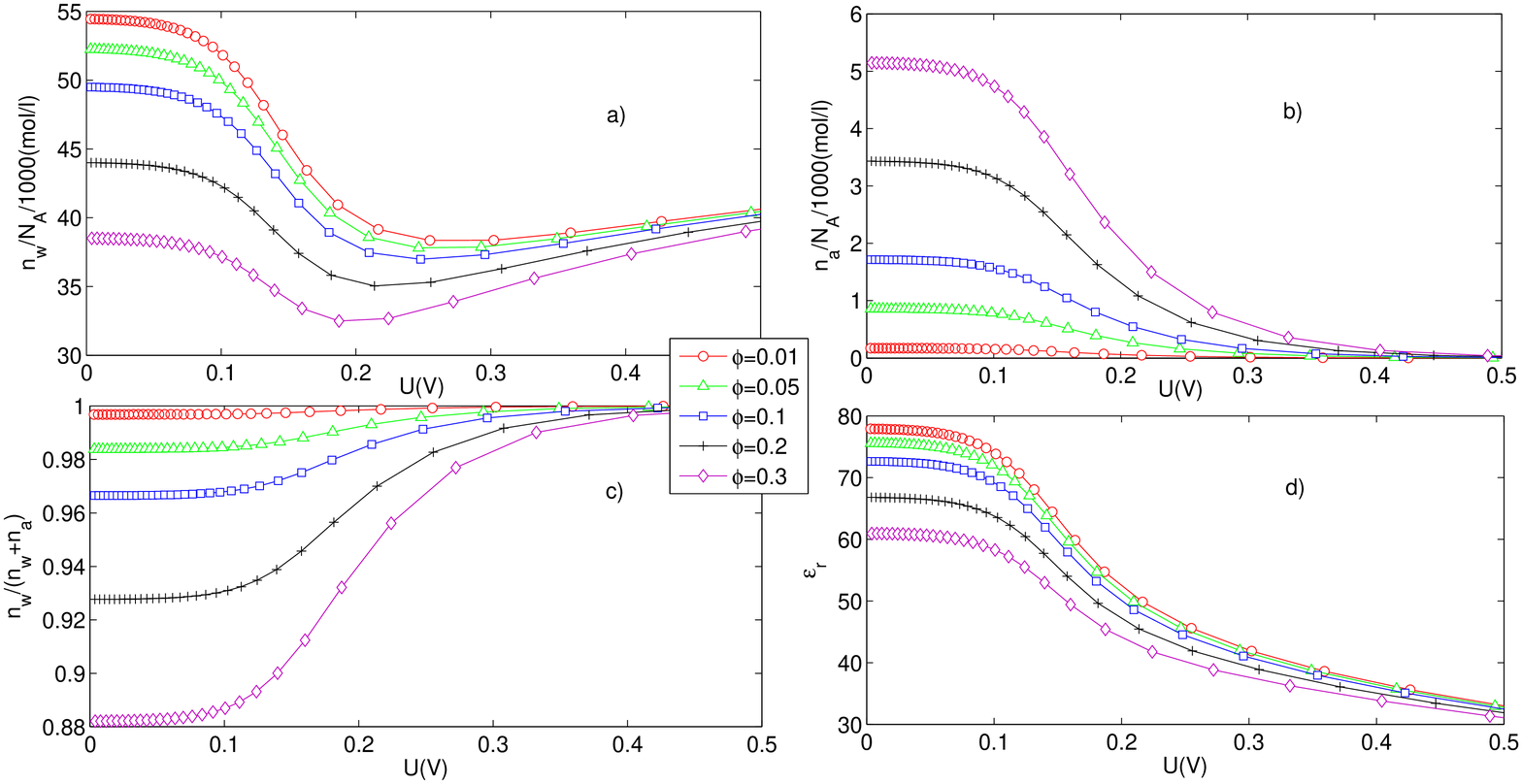}
\caption{(Color online) At the charged surface, the number density of water molecules (a), the number density of ethylalcohol molecules (b), the portion of water molecules in solvent mixtures (c) and the permittivity (d) as a function of the surface voltage.  The parameters are the same as in Fig. \ref{fig:1}.}
\label{fig:2}
\end{figure}

Fig. \ref{fig:2}(a) shows the number density of water molecules at the charged surface. As one can see, for all the cases, the profile is non-monotonous with respect to the surface voltage.  Like in a pure aqueous electrolyte($\phi=0$) shown in \cite%
{Gongadze_genphysiol_2013, Sin_electrochimica_2015, Sin_pccp_2016}, for all the binary solvent compositions, the number density of water molecules has a valley.  For a high binary solvent composition($\phi=0.3$), the number density of water molecules at high surface voltages is higher than for zero surface voltage. The fact is a direct result of strong exclusion of ethylalcohol molecules. It should be also noted that the higher the surface voltage, the smaller the difference in number density of water molecules between different binary solvent compositions

Fig. \ref{fig:2}(b) shows the number density of ethylalcohol molecules at the charged surface. Unlike the number density of water molecules, the number density of ethylalcohol molecules monotonously decreases with increasing the surface voltage. This means that ethylalcohol molecules are continuously excluded according to the increase in surface voltage. For every binary solvent composition, at a surface voltage above 0.3V, the number density of ethylalcohol molecules nearly disappears in the vicinity of the charged surface.

Fig. \ref{fig:2}(c) shows the portion of water molecules to both solvent molecules at the charged surface according to surface voltage. Increase in the surface voltage causes the portion of water molecules to increase towards saturation value of 1. At high surface charge densities, the difference between the values for different binary solvent compositions disappears by the same reason as in Fig. \ref{fig:2}(a) and Fig. \ref{fig:2}(b).

Fig. \ref{fig:2}(d) shows the spatial dependence of relative permittivity according to the surface voltage. 
For every binary solvent composition the permittivity decreases with increasing the surface voltage. The decrease is due to the accumulation of the counterions and exclusion of solvent molecules near the charged surface.  It is shown that the higher the binary solvent composition, the lower the permittivity of binary solvent mixture. This behavior is proved from Eq.(\ref{eq:18}) using the fact that the dipole moment of an ethylalcohol molecule is lower than for a water molecule.

\begin{figure}
\includegraphics[width=0.5\textwidth]{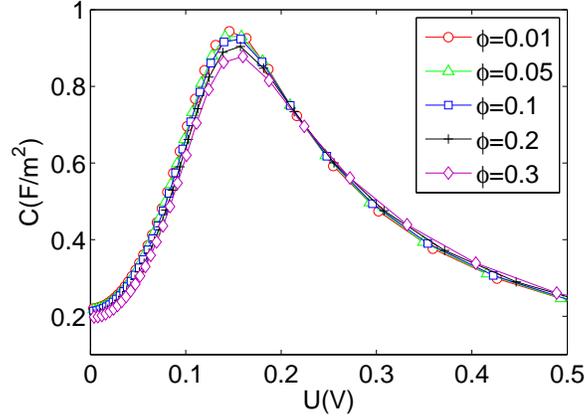}
\caption{(Color online) Differential  capacitance  as  a  function  of  surface  voltage  for different binary solvent composition.
All parameters are the same as in Fig. \ref{fig:2}.}
\label{fig:3}
\end{figure}

\begin{figure}
\begin{center}
\includegraphics[width=1\textwidth]{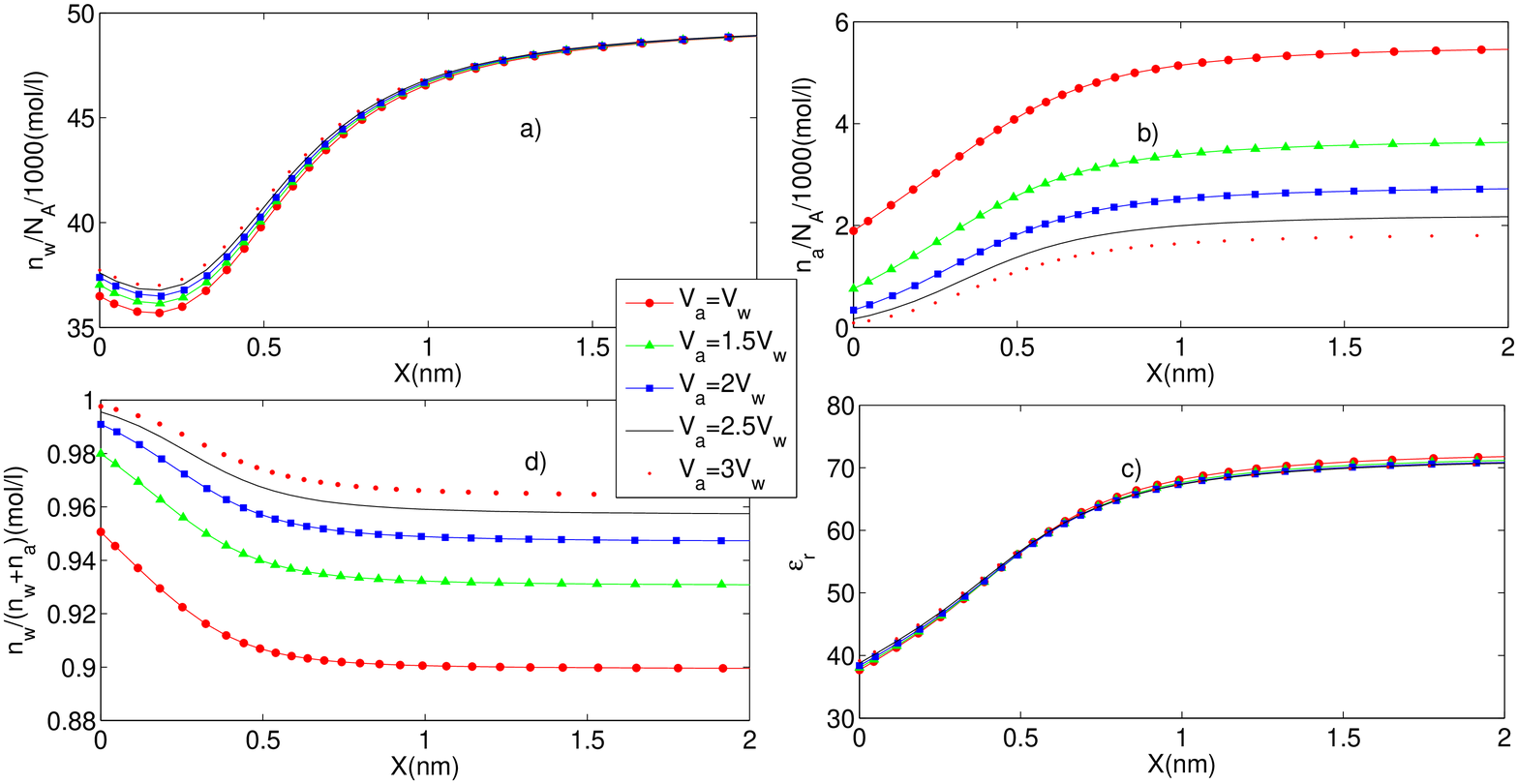}
\caption{(Color online) The number density of water molecules (a), the number density of added solvent molecules (b), the portion of water molecules in solvent mixtures (c) and the permittivity (d) as a function of the distance from the charged surface in binary solvent mixtures for different volumes of an added solvent molecule. The binary solvent composition, the dipole moment of an added solvent molecule and the surface charge density are 0.1, 0.5D and $0.2C/m^2$, respectively. Other parameters are the same as in Fig. \ref{fig:2}.}
\label{fig:4}
\end{center}
\end{figure}
Fig. \ref{fig:3} demonstrates the differential capacitance curve according to the surface voltage for different binary solvent composition. It should be noted that increasing binary solvent composition causes maximum capacitance to  slightly lower. The fact is attributed to decrease in surface charge density due to decrease of permittivity according to the binary solvent composition as shown in Fig. \ref{fig:2}(d).

Fig. \ref{fig:4}(a) and Fig. \ref{fig:4}(b) show the number densities of water molecules and added solvent molecules according to the distance from the charged surface for the cases having different added solvent size, respectively. As shown in the Fig. \ref{fig:4}(a) and Fig. \ref{fig:4}(b),  for a smaller volume of an added solvent molecule water molecules  are excluded more strongly near the charged surface, whereas added solvent molecules are excluded more weakly than for the original case. This fact is clearly demonstrated from Eq. (\ref{eq:101}), which suggests that the number density of added solvent having a large volume is smaller than for the added solvent with a smaller volume.

	Fig. \ref{fig:4}(c) demonstrates the portion of water molecules to the number density of both solvent molecules according to the distance from the charged surface. For every added solvent volume, the portion decreases with increasing the distance from the charged surface.  Furthermore, a large volume of an added solvent molecule corresponds to a small value of portion of added solvent molecules near the charged surface compared to the case of a smaller volume of added solvent molecules. These facts are proved by combining the results of Fig. \ref{fig:4}(a) and Fig. \ref{fig:4}(b).

	Fig. \ref{fig:4}(d) shows the permittivity of binary solvent mixture near the charged surface.    It is interesting that there is no difference in permittivity between different volumes of an added solvent. This is a direct result from the fact that for a small volume of an added solvent the change of permittivity due to decrease in number density of water molecules is compensated by an increase in number density of added solvents compared to the case having a larger volume.

\begin{figure}
\begin{center}
\includegraphics[width=1\textwidth]{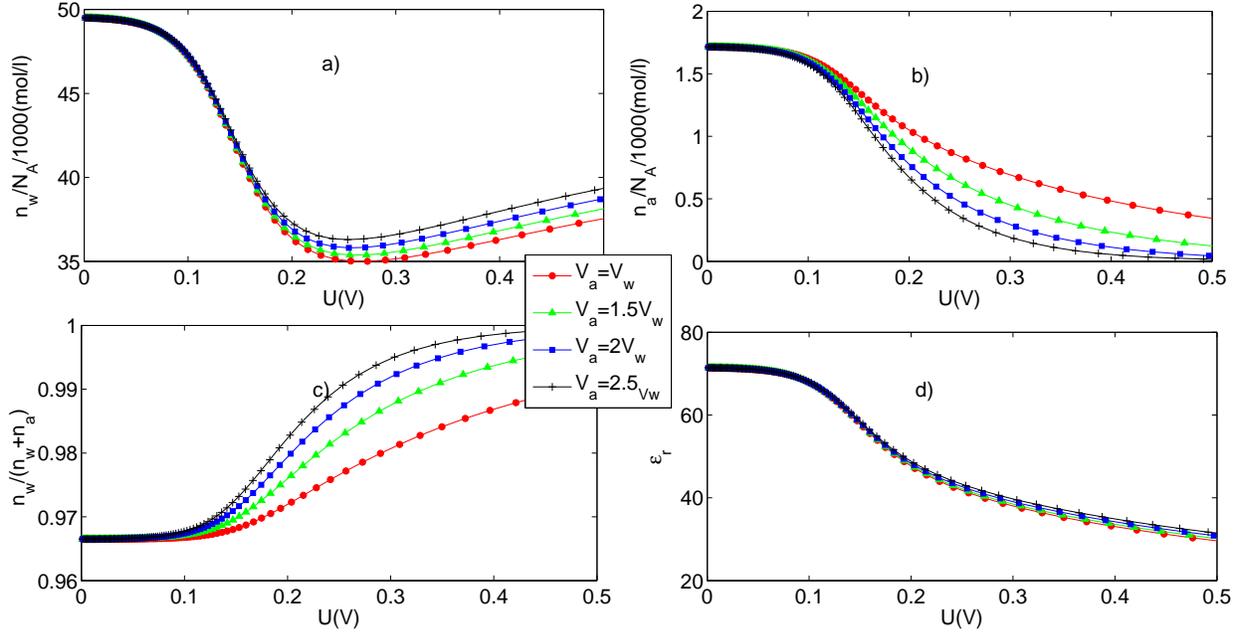}
\caption{(Color online) At the charged surface, the number density of water molecules (a), the number density of added solvent molecules (b), the portion of water molecules in solvent mixtures (c) and the permittivity (d) as a function of the surface voltage in binary solvent mixtures for different volumes of an added solvent molecule. All the parameters are the same as in Fig. \ref{fig:4}.}
\label{fig:5}
\end{center}
\end{figure}

Fig. \ref{fig:5}(a) and Fig. \ref{fig:5}(b) show the number densities of water molecules and added solvent molecules at the charged surface for different volumes of an added solvent molecule, respectively. As in Fig. \ref{fig:2}(a), the profile of water molecules has a non-monotonous behavior with respect to the surface voltage. Fig. \ref{fig:5}(a) and Fig. \ref{fig:5}(b) indicate that the higher the surface voltage, the larger the difference in number density of solvent molecules between the cases for different volumes of an added solvent molecule. This means that at high surface voltages added solvent molecules are more strongly excluded from the charged surface compared to water molecules.  The fact is also demonstrated from Eq. \ref{eq:101} which implies that when increasing the surface voltage, the number density of added solvent molecules with a larger volume decreases more sharply compared to the case of original volume.

Fig. \ref{fig:5}(c) and Fig. \ref{fig:5}(d) show the portion of water molecules to both solvent molecules and the permittivity of binary solvent mixture at the charged surface according to surface voltage, respectively. In the region of high surface voltages, for both cases of portion of water molecules and permittivity, the difference between the corresponding values for different volumes of an added solvent molecule is enhanced by the same reason as in Fig. \ref{fig:5}(a) and Fig. \ref{fig:5}(b).

  \begin{figure}
\begin{center}
\includegraphics[width=0.5\textwidth]{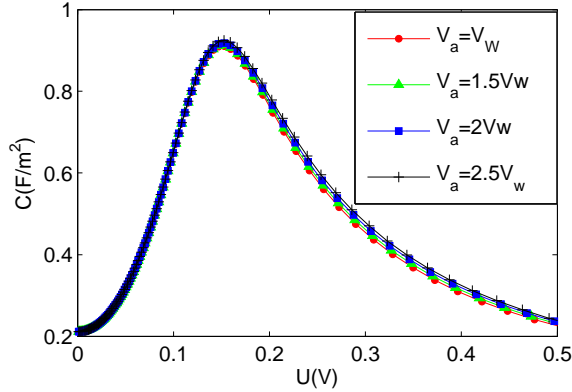}
\caption{(Color online) The differential capacitance as a function of surface voltage for our approach($V_{-}= 0.33nm^3, V_{+}= 0.15nm^3$) and Gongadze-Iglic model($V_{-}= V_{+}= V_{w}$). Other parameters are the same as in Fig. \ref{fig:4}.}
\label{fig:6}
\end{center}
\end{figure}

Fig. \ref{fig:6} demonstrates the differential capacitance curve according to the surface voltage for different volumes of an added solvent molecule. When increasing an added solvent molecular size, not only maximum capacitance but also capacitances at high surface voltages slightly increase. In the same way as in Fig. \ref{fig:3}, this is attributed to the increase in the permittivity shown in Fig. \ref{fig:5}(d).

\begin{figure}
\begin{center}
\includegraphics[width=1\textwidth]{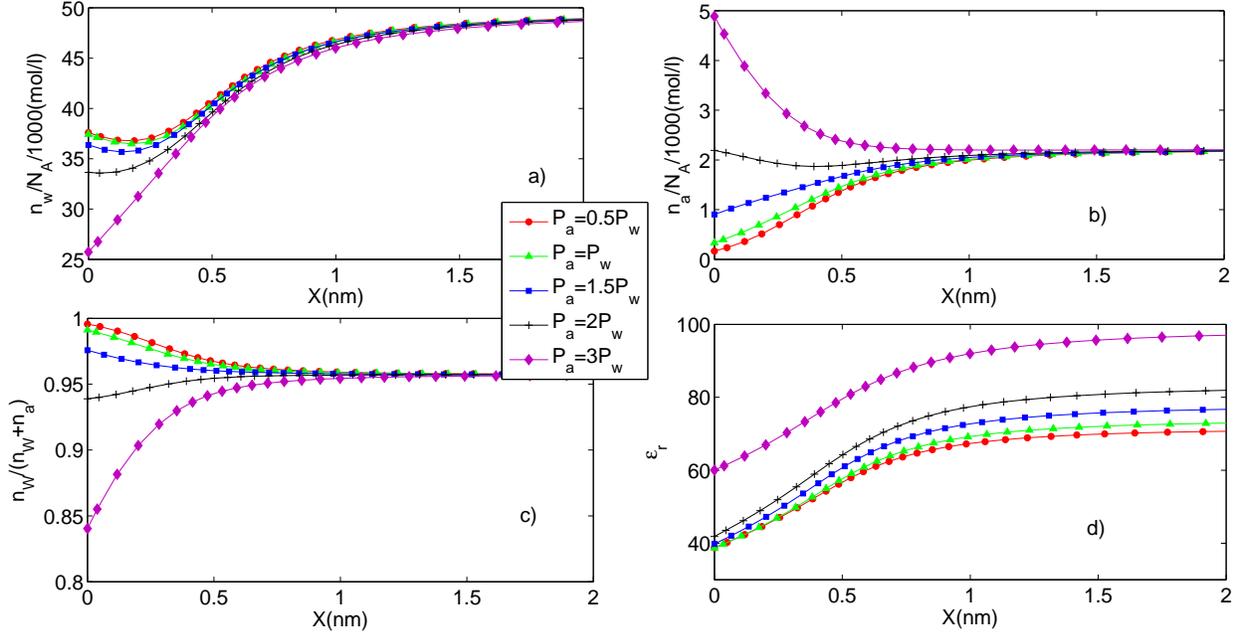}
\caption{(Color online) The number density of water molecules (a), the number density of added solvent molecules (b), the portion of water molecules in solvent mixtures (c) and the permittivity (d) as a function of the distance from the charged surface in binary solvent mixtures. The binary solvent composition, the volume of an added solvent molecule and the surface charge density are 0.1, 2.5$V_w$ and $0.2C/m^2$, respectively. Other parameters are the same as in Fig. \ref{fig:1}.}
\label{fig:7}
\end{center}
\end{figure}

Fig. \ref{fig:7}(a) and Fig. \ref{fig:7}(b) show the spatial dependences of the number densities of water molecules and added solvent molecules near a charged surface with surface charge density $\sigma=0.2C/m^2$. Circles, Triangles, Squares , Plus Signs and Diamonds stand for the number densities of water molecules for the cases having the dipole moment of an added solvent molecule $p_a/p_w$=0.5, 1, 1.5, 2.0, 3.0 respectively.  Fig. \ref{fig:7}(a) and Fig. \ref{fig:7}(b) imply that a large dipole moment of an added solvent molecule causes water molecules to strongly exclude, added solvent molecules to weakly exclude near the charged surface. This fact is attributed to the fact that term $\frac{\sinh\left(\gamma p_w \beta E\right)}{\left(\gamma p_w \beta E\right)}$  of Eq. (\ref{eq:101})  increases with increasing the dipole moment of solvent.
	
Fig. \ref{fig:7}(c) shows the portion of water molecules to both solvent molecules according to the distance from the charged surface. It is shown that a large dipole moment of added solvent molecules corresponds to a small value of portion of added solvent molecules near the charged surface compared to the case of a smaller dipole moment. This fact is a direct result of the consequences of Fig. \ref{fig:7}(a) and Fig. \ref{fig:7}(b).

	Fig. \ref{fig:7}(d) shows the permittivity of binary solvent mixture near the charged surface.   It is shown that for a large dipole moment of an added solvent molecule, the permittivity of binary solvent mixture is larger than for a smaller dipole moment.  This can be understood from the Eq. (\ref{eq:18}) , which implies that the larger the dipole moment of water molecules, the larger the permittivity of binary solvent mixture.

\begin{figure}
\begin{center}
\includegraphics[width=1\textwidth]{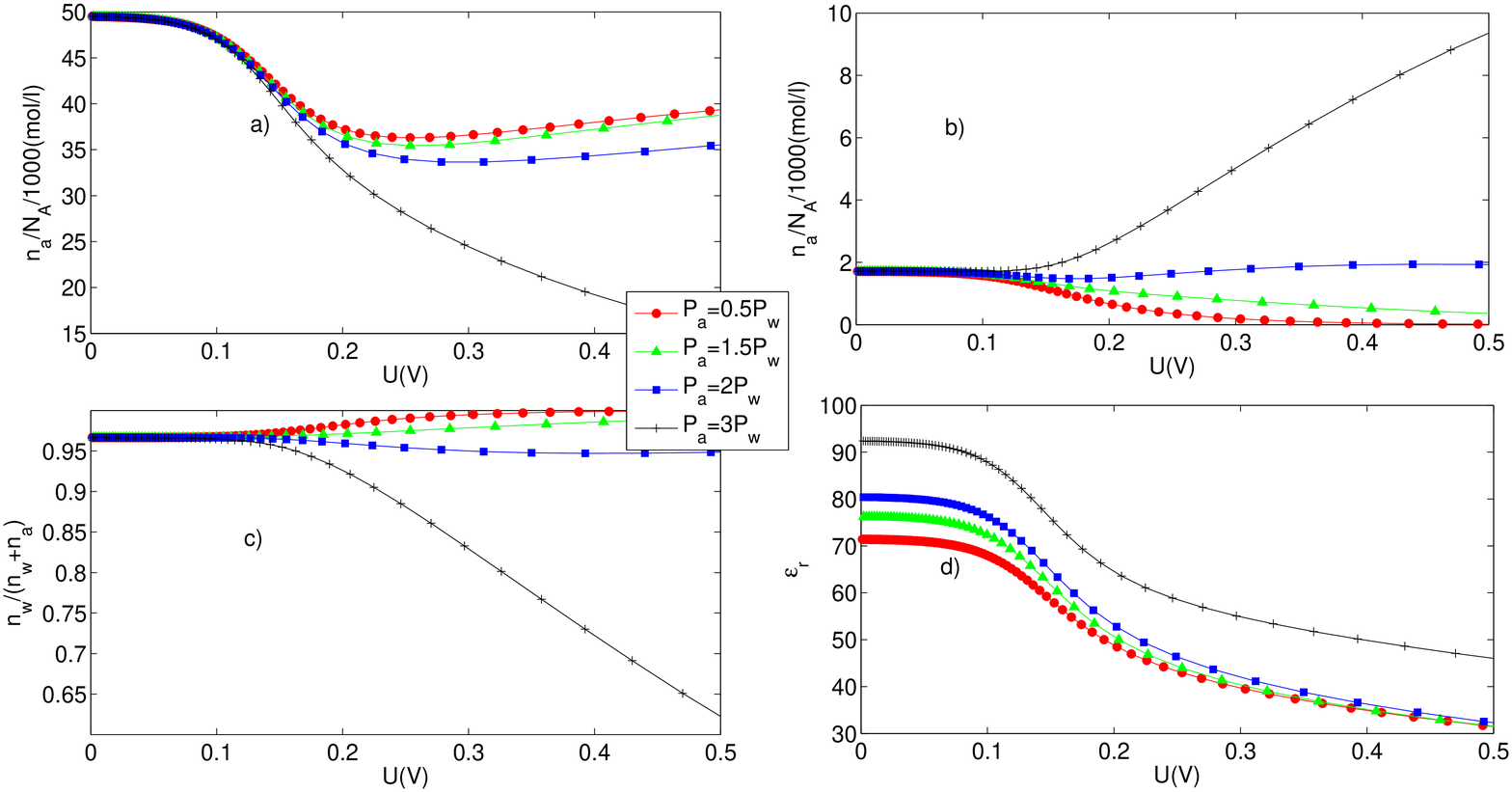}
\caption{(Color online)  At the charged surface, the number density of water molecules (a), the number density of added solvent molecules (b), the portion of water molecules in solvent mixtures (c) and the permittivity (d) as a function of the surface charge density in binary solvent mixtures. All the parameters are the same as in Fig. 7.}
\label{fig:8}
\end{center}
\end{figure}
Fig. \ref{fig:8}(a) and Fig. \ref{fig:8}(b) show the number densities of water molecules and added solvent molecules at the charged surface for different dipole moments of an added solvent molecule, respectively. Fig. \ref{fig:8}(a) and Fig. \ref{fig:8}(b) indicate that the higher the surface voltage, the larger the difference in number density of solvent molecules between the cases for different dipole moments of an added solvent molecule. 
More importantly, it is noticeable that the increase in the dipole moment of an added solvent molecule lowers the number density of water molecules and causes an increase in the number density of added solvent molecules. In particular, in the case of $p_a=3p_w$ at the charged surface, the number density of added solvent molecules increases with increasing the surface voltage and the water molecules are strongly excluded from the vicinity of charged surface. This is a result of Eq. (\ref{eq:101}), which implies a large density corresponding to  a large dipole moment of a solvent molecule.

Fig. \ref{fig:8}(c) shows the portion of water molecules to both solvent molecules at the charged surface. It is shown that for the case of a small dipole moment of an added solvent molecule $p_a=1,1.5,2p_w$, the portion of water molecules increases with increasing the surface voltage, whereas in the case of $p_a=3p_w$, the portion rapidly decreases. This should be explained as follows.
In \cite%
{Li_pre_2011, Li_pre_2012}, it was demonstrated that the key factor for the stratification of counterions is $Z/V$, the ratio of valency of counterions to the volume of counterions.  Recently, in searching to well understand this result using a mean-field approach \cite%
{Sin_electrochimica_2016}, we quite strictly proved that when orientational ordering of water dipoles is considered, the ratio still plays the key factor for counterions stratification and orientational ordering of water dipoles enhances the phenomena. In a similar way,  we intuitively feel that the key factor for solvent competition will be defined as the ratio of the dipole moment to the volume of a solvent molecule, $p/V$. In other word, the portion of the solvents having a larger value of the key factor increases with decreasing the distance from the charged surface, whereas the portion of other solvents decreases. Indeed, the idea can be easily deduced from the Eqs. (\ref{eq:10}, \ref{eq:101}).  Moreover, this criterion is consistent with all the results above mentioned for the number densities of solvents and the portion of solvents.
\begin{figure}
\begin{center}
\includegraphics[width=0.5\textwidth]{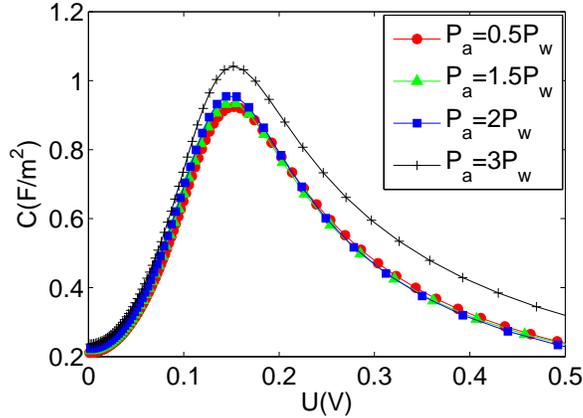}
\caption{(Color online) Differential  capacitance  as  a  function  of  surface  voltage  for different dipole moments of second solvent. All the parameters are the same as in Fig. 7.}
\label{fig:9}
\end{center}
\end{figure}
	Fig. \ref{fig:8}(d) shows the permittivity at the charged surface according to the surface voltage. It indicates that the larger the dipole moment of added solvent molecules, the larger the permittivity at the charged surface. 
Although there are small differences in permittivity between the cases of $p_a=0.5,1.5,2p_w$, for the case of $p_a=3p_w$, the permittivity is significantly high compared to other cases. These facts are directly obtained from permittivity formula and the concept of solvent competition.

	Fig. \ref{fig:9} shows the differential capacitance curve according to the surface voltage for different dipole moments of an added solvent molecule. In the same way as in Fig. 6, the maximum capacitances for all the cases are in the same order as in permittivity shown in Fig. \ref{fig:8}(d). In the case of $p_a=3p_w$, the values are quite differs from ones for other cases and very similar to for the binary solvent composition of 1. This is another demonstration of competition between water molecules and added solvent molecules.

In fact, the authors of \cite%
{Kashyap_chemphys_2007} also took into account the differences in solvent diameters, dipoles moments, and ionic size using the mean spherical approximation. However, their study concentrated on the role of partial solvent polarization densities around an ion in a completely asymmetric binary dipolar mixture, but did not treat the phenomena relevant to asymmetric depletion between two kinds of solvent.

Although our approach and results are original, it is required that the approach is extended to be treated important issues such as preferential solvation and short-range interaction between water and added solvent. In particular, such improvement is very important for the interaction between charged membranes\cite%
{Andelman_jphyschemb_2009}. In addition, to confirm the validity of our results, it is necessary that more complete study such as Monte Carlo or molecular dynamics simulations will be performed.

\section{Conclusion}

In this work, the differences in size and dipole moment between solvent molecules are described by updating the mean-field model \cite%
{Sin_electrochimica_2016} which accounts for non-uniform size effects and orientational ordering of water dipoles. 
Our model predicts that the ratio of dipole moment of solvents to the volume of solvents p/V plays the role of key factor for competition between solvents. In other words, when the key factor of a solvent molecule is large compared to the other, the portion of the solvent increases with respect to not only the distance from the charged surface but also the surface voltage. 
We conclude that when the key factor for added solvent is smaller than for water, the differences in volume and dipole moments of added solvents hardly affect the properties of binary solvent mixture such as permittivity and differential capacitance. 
However, if the key factor for the added solvent is larger than for water, the permittivity and differential capacitance are dramatically changed in the same way as in pure added solvent solution. 
The influence of  binary solvent composition, volume and dipole moment of solvent molecules  on  relative  permittivity  and  on  differential  capacitance  of  EDL  in  binary  solvent mixtures  has  been reported and discussed in more detail.
In  practice,  our  modified  approach  and  results  can  be  used  to describe  the  phenomena in binary solvent mixtures such as transition to a condensed phase of DNA \cite%
{Tsori_pnas_2007, Andelman_jphyschemb_2009} and  differential  capacitance  of  electric  double  layer  capacitor \cite%
{Chen_jphyschemc_2015, quiroga_jelechemsoc_2014, Wang_physchem_2009}.

\end{document}